\documentclass[11pt, a4paper]{article}
\usepackage{inputenc}
\usepackage{graphicx}

\title{Persistent currents, deformation and collectivity in the
  many-boson yrast problem on the circle}

\author{E. J. V. de Passos\footnote{passos@if.usp.br} and A. F. R. de
  Toledo Piza\footnote{piza@if.usp.br} \\ Instituto de F\'{i}sica,
  Universidade de S\~{a}o Paulo, S\~{a}o Paulo, Brazil}

\begin{document}

\maketitle

\begin{abstract}

  Properties of the yrast states of a system of $N$ bosons confined to
  a one-dimensional ring and interacting via contact forces is
  examined both variationally and by numerical diagonalizations. The
  latter allow for obtaining numerical correlated many-body
  wave functions explicitly. The study of correlation functions
  involving different yrast states indicates that a quantum phase
  transition previously detected in the properties of the ground state
  in the case of attractive two-body interactions is an yrast
  phenomenon involving the onset of `deformation', in the sense given
  to this term by Bohr and Mottelson in connection with the
  description of nuclear spectra, including enhanced transition
  operators and emergence of a shared intrinsic correlation
  structure. In this case the moment of inertia of the deformed state
  is essentially the rigid moment of inertia, `intrinsic' states being
  essentially degenerate.

\end{abstract}

\section{Introduction.} 

Persistent matter currents in the multiply connected ring geometry
have now been experimentally realized and studied many times in
connection with ultra cold bosonic gases\cite{exp,exphslip,expLargeL}
in order to explore super fluid properties of the underlying many-body
system. The most salient feature present in all cases is the
`quantization of the persistent current with macroscopic values of the
angular momentum' which was dubbed {\it macroscopic isomerism} by Bohr
and Mottelson more than 50 years ago\cite{BoMot}, in connection with
the quantization and (meta-)stability of electric currents in
superconductors. In the case of the bosonic gases, the relevant
`macroscopic' values of the angular momentum $L$ are (in units of
$\hbar$) the integer multiples of the number of particles $N$.

Such macroscopic quantization was long ago dealt with, in the Bose gas
context, by F. Bloch\cite{Bloch}. This work appeared in the wake of
related work on superconductivity\cite{BloJoseph} and is based on a
detailed analysis of the minimum energies involved in imparting given
amounts of angular momentum to the system ground state. The analysis
was carried out in the context of a simple (one-dimensional)
microscopic quantum mechanical model, revived and extended more
recently by several authors\cite{Rina,var,list}. In particular, the
occurrence of a quantum phase transition was pointed out in \cite{Rina}
in a study of the ground state properties with attractive two-body
effective interactions. It was signalled by the `solitonic' breaking of
the rotational symmetry in an effective mean field approach, and
corroborated by the behavior of the ground state two-body correlation
function obtained by diagonalizing the effective Hamiltonian in a Fock
subspace with good particle number and total angular momentum.

Symmetries, such as rotational invariance, play a most important role
in organizing the dynamical properties of finite quantum many-body
systems. However, is also well known that there are many
phenomenologically salient features which develop across the cleavage
resulting from them. A familiar example of this is the occurrence of
rotational spectra in molecules and atomic nuclei which involve
subspaces of different angular momenta, even though they are
dynamically disjoint as a result of rotational invariance. Such
features are currently described in terms of the stability of certain
symmetry violating correlation properties, or `deformations'\cite{BM}.
The resulting syndrome is in fact the finite-system counterpart of the
spontaneous breaking of symmetries which can occur in extended
systems.

In this work we study, as a function of angular momentum, the low
energy quantal spectrum of a dilute bosonic gas confined to a quasi
one-dimensional toroidal trap, assuming the usual zero-range effective
two-body interaction based on the s-wave scattering length. This is
done taking advantage of the strictly {\it finite} nature of the
system and adopting two independent approaches, namely a) using a
variational treatment involving product states with possibly broken
rotational symmetry while constraining the mean value of the angular
momentum per particle and b) by carrying out numerical diagonalization
of the Hamiltonian in many-body subspaces of definite particle number
and angular momentum, built from the different possible occupations of
a restricted set of relevant single-particle states. Both effectively
repulsive and attractive two-body interactions are considered. We find
in both cases that the qualitative features of the system are at most
weakly dependent on the number of particles provided the effective two
body interaction strength is scaled with $1/N$ so as to preserve the
relative weight of the kinetic (one-body) and the interaction
(two-body) dynamical ingredients at fixed trap geometry.

We find that the effect of repulsive effective two-body interactions
is essential for providing conditions for macroscopic isomerism (or
persistent current meta-stability) in the model\cite{Bloch,leggett}.
The strengths prevailing in current realistic experimental situations
are in fact orders of magnitude larger as compared with threshold
values obtained for several lowest values of $L/N$. Attractive
two-body effective interactions are found to lead to the onset of a
regime in which the excitation energies of the lowest states of each
successive value of the total angular momentum correspond to the
energies of an essentially rigid rotational band with the rigid moment
of inertia. In this way, the quantum phase transition pointed out in
ref. \cite{Rina} appears as an yrast phenomenon associated to the
emergence of a correlation structure common to a family of states with
different values on the total angular momentum. This is clearly akin
to `deformation' as found in connection with rotational spectra in
nuclear and molecular physics.  We find also that the correlation
structure induced by attractive (repulsive) two-body interactions
strongly enhances (appreciably reduces) transition matrix elements
involving a one body angular momentum exchange operator along the
yrast line and so will affect decay rates for processes dependent on
such matrix elements. It should be kept in mind, however that {\it
  observed} decay probabilities (as e.g. in \cite{exphslip}) may well
involve degrees of freedom beyond those of the present quasi
one-dimensional model.

\section{Hamiltonian.} 

The Hamiltonian for scalar bosons in a tight toroidal trap of radius
$R$ and cross section $S$ is
 
\begin{equation}
H= \frac{\hbar^{2}}{2 M R^{2}} \int_{0}^{2 \pi} d\varphi \psi^{\dagger}
(\varphi) l_{z}^{2} \psi (\varphi) +\frac{1}{2} \frac{U_{c}}{RS}
\int_{0}^{2 \pi} \psi^{\dagger} (\varphi) \psi^{\dagger} (\varphi) \psi
(\varphi) \psi (\varphi) d\varphi
\label{Hfield}
\end{equation}

\noindent where $ \psi (\varphi),\psi^{\dagger} (\varphi) $ are the
field operators, $l_{z}= -i\frac{\partial}{\partial \varphi} $ is the
angular momentum operator in units of $\hbar$ and $U_{c}=4 \pi
\frac{\hbar^2 a}{M} $ is the strength of the effective two-body, 3-D
contact interaction, $a$ being the scattering length.

Introducing bosonic creation and annihilation operators
$b_{m}^{\dagger}$ and $b_{m}$ for atoms in the single particle angular
momentum eigenstates

\begin{equation}
\phi_{m}(\varphi)=\frac{1}{\sqrt{2 \pi}} e^{im\varphi},\hspace{1cm}
\int_{0}^{2\pi} d\varphi \phi_{m^{\prime}}^*(\varphi)\phi_{m}(\varphi)=
\delta_{m^{\prime },m},
\label{strbase}
\end{equation}

\noindent and expanding the field operators as         

\begin{equation}
\psi(\varphi)=\sum_m\phi_m(\varphi)b_m,
\label{straight}
\end{equation}

\noindent the Hamiltonian (\ref{Hfield}) can be expressed as

\begin{eqnarray}
H&=&\frac{\hbar^{2}}{2 M R^{2}}\sum_{m} m^{2} b_{m}^{\dagger}b_{m}+
\frac{\Lambda}{2}\sum_{m_i} b_{m_{1}}^{\dagger}b_{m_{2}}^{\dagger}
b_{m_{3}} b_{m_{4}}\delta_{m_{1}+m_{2},m_{3}+m_{4}}\equiv\nonumber\\
&\equiv& K+V
\label{Hbase}
\end{eqnarray}

\noindent where $\frac{\hbar^{2}}{2 M R^{2}}$ and $\Lambda\equiv
\frac{U_{c}}{2\pi RS}$ set the kinetic and interaction energy scales.
                                                                  
This second quantized Hamiltonian commutes with the number operator
${\cal{N}}=\sum_mb^\dagger_mb_m$. In each of the Fock space sectors
characterized by the particle number $N$ (where ${\cal{N}}$ acts as
the corresponding multiple of the unit operator), it can be split
into `center of mass' and `intrinsic' parts by introducing the total
angular momentum operator (in units of $\hbar$) $L_z\equiv\sum_m
m\;b_m^\dagger b_m$, also a constant of motion, and writing its
kinetic energy part as

\[
K=\frac{\hbar^2L_z^2}{2NMR^2}+K_{\rm int}
\]

\noindent  with

\[
K_{\rm int}=\frac{\hbar^2}{4NMR^2}\sum_{m_1m_2}
(m_1-m_2)^2 b_{m_{1}}^{\dagger}b_{m_{2}}^{\dagger}b_{m_{2}} b_{m_{1}}
\]

\noindent which can be easily verified to be algebraically equivalent
to the kinetic energy term of eq. (\ref{Hbase}). The center of mass
part is clearly just the rigid rotational energy $\frac{\hbar^2L_z^2}
{2NMR^2}$, which commutes both with the intrinsic part $K_{\rm int}$
of the kinetic energy and with the potential energy term $V$. One has
therefore 

\begin{equation}
H=K_{\rm CM}+H_{\rm int}\hspace{1cm}{\rm with}\hspace{1cm}
H_{\rm int}=K_{\rm int}+V.
\label{Hsplitt}
\end{equation}

\noindent Since the total angular momentum $L_z$ is also a constant of
motion, the stationary states of $H$ and of $H_{\rm int}$ can be
chosen as states with good total angular momentum $L$.

\section{The spectrum of $H_{\rm int}$.} 

Here we briefly review the results of F. Bloch \cite{Bloch} within the
specific context of the Hamiltonian (\ref{Hfield}) or (\ref{Hbase}). In
view of the decomposition (\ref{Hsplitt}) of this Hamiltonian, the
eigenvalues $E_L$ of states with total angular momentum $L$ can in
turn be split as

\begin{equation}
E_L=\frac{\hbar^2L^2}{2NMR^2}+e_{\rm int}^{(L)}
\label{blodec}
\end{equation}

\noindent where the last term is an eigenvalue of the intrinsic
Hamiltonian $H_{\rm int}$ also associated with the total angular
momentum $L$. The corresponding $N$-particle eigenstate can be
expanded in the occupation number base vectors $|\{n_m\}_{N,L}\rangle
\equiv\prod_m\frac{1}{\sqrt{n_m!}}\left(b^\dagger_m\right)^{n_m}
|0\rangle$, built on single particle states with good angular momentum
$m$ and satisfying the conditions $\sum_m n_m=N$ and $\sum_m
m\,n_m=L$, as

\begin{equation}
|\Psi_{E_L^{(j)}}\rangle=\sum_{\{n_m\}_{N,L}}C_{\{n_m\}_{N,L}}^{(j)}
\prod_m\frac{1}{\sqrt{n_m!}}\left(b^\dagger_m\right)^{n_m}|0\rangle.
\label{mbstate}
\end{equation}

\noindent Since the intrinsic Hamiltonian $H_{\rm int}$ depends only
on relative angular momenta of particle pairs\cite{note}, it follows
that the vectors obtained from this expansion just by shifting all the
single particle angular momenta $m$ by an integer $l$, i.e.

\[
|\Psi_{E_{L+l N}^{(j)}}\rangle=\sum_{\{n_m\}_{N,L}}C_{\{n_m\}_{N,L}}^{(j)}
\prod_m\frac{1}{\sqrt{n_m!}}\left(b^\dagger_{(m+l)}\right)^{n_m}
|0\rangle,
\]

\noindent will have the same intrinsic energy, while the total angular
momentum is shifted by $l N$, so that the center of mass energy is
$\frac{\hbar^2(L+l N)^2}{2NMR^2}$. This implies that the spectrum of
$H_{\rm int}$ is periodic in $L$ with period $N$. Furthermore,
invariance of the spectrum under the replacement of the base vectors
$|\{n_m\}_{N,L}\rangle$ by $|\{n_{-m}\}_{N,-L} \rangle$ guarantees
that the $L$-dependence of the periodic intrinsic spectrum is also
reflection symmetric.

Of special relevance here is the set lowest eigen-energies for each
value of the total angular momentum (the set of the so called {\it
  yrast} states, or the {\it yrast line}\cite{Mottelson}). In view of
the above properties, the energy eigenvalues on this line will be
given as the addition of the set of lowest intrinsic energies for
increasing values of the total angular momentum $L$ (which is periodic
in $L$ with period $N$) to the the center of mass energy parabola
$\frac{\hbar^2L^2} {2NMR^2}$. As argued in ref. \cite{Bloch} (see also
ref. \cite{leggett}), the occurrence of minima along the yrast line at
non vanishing values of the total angular momentum signals the
existence of meta-stable states with persistent currents in the system
when additional couplings allowing for angular momentum and energy
exchange between the system and the constraining environment.

\section{Characterization of the yrast line I - variational
approach.}  

In view of the general periodicity features just described, it is
convenient to write the total angular momentum $L$ (in units of
$\hbar$) as $L=N(l+\nu)$, $l$ being the integer part of the ratio
$L/N$ and $\nu=0, \frac{1}{N}, \frac{2}{N},\dots, \frac{N-1}{N}$. In
this way, it is easily ascertained that the intrinsic energies for a
system of {\it free} bosons in the ring are given by 

\[
\frac{e_{\rm int}^{(L)}}{N}=\frac{\hbar^2}{2MR^2}\,\nu(1-\nu),
\hspace{1cm}\nu=0,\frac{1}{N},\frac{2}{N},\dots,\frac{N-1}{N}
\]

\noindent They are independent of $l$, consistently with the
periodicity of the intrinsic spectrum. One sees moreover i) that
the {\it intrinsic} yrast energies display cusp minima at the integer
values of $L$, and ii) that the only minimum of the yrast line
(including the center of mass rotational energy) for the free bosons
occurs at $L=0$.

In order to explore variationally the effects of the effective two
body interaction on the yrast energies we make use of the family of
`condensate' states\cite{var}

\[
|\Psi\rangle\equiv\frac{1}{\sqrt{N!}}\left(A^\dagger\right)^N|0
\rangle
\]

\noindent with

\begin{equation} 
A^\dagger\equiv c_{l-1}b_{l-1}^\dagger+c_lb_l^\dagger+c_{l+1}
b_{l+1}^\dagger\hspace{0.5cm}{\rm and}\hspace{.5cm}|c_{l-1}|^2+
|c_l|^2+|c_{l+1}|^2=1,
\label{constr1}
\end{equation}

\noindent so that $[A,A^\dagger]=1$. These states are used to obtain
approximations to the yrast line energies associated with angular
momenta $L=N(l+\nu)$. The approximate energies are given by the value
of the energy functional $\langle\Psi|H|\Psi \rangle$ in the state
$|\Psi\rangle$ which minimizes this value under the constraint
$\langle\Psi|L_z|\Psi\rangle=L$, which is expressed in terms of the
amplitudes involved in the definition of $A^\dagger$ as

\begin{equation}
|c_{l+1}|^2-|c_{l-1}|^2=\nu.
\label{constr2}
\end{equation}

\noindent The constraints (\ref{constr1}) and (\ref{constr2}) are
taken into account by introducing the parameterization

\[
c_l=e^{i\alpha_l}|c_l|,\;\,\;\,c_{l+1}=e^{i\alpha_{l+1}}\sqrt
{\frac{1+\nu-|c_l|^2}{2}},\;\,\;\,c_{l-1}=e^{i\alpha_{l-1}}\sqrt
{\frac{1-\nu-|c_l|^2}{2}}
\]

\noindent in which the variational parameters are $|c_l|$ (restricted
to the domain $0<|c_l|^2<1-\nu$) and the phases $\alpha_{l-1}$,
$\alpha_l$ and $\alpha_{l+1}$. The evaluation of the energy functional
is then straightforward. In units of $\frac{\hbar^2}{2MR^2}$ the
energy functional per particle reads

\begin{eqnarray}
&&{\cal{F}}_l(|c_l|,\Phi,\nu)\equiv\frac{2MR^2}{\hbar^2}
\frac{\langle\Psi|H|\Psi\rangle}{N}=\frac{L^2}
{N^2}+1-\nu^2-|c_l|^2+\\ &+&g\left(\frac{3}{4}+\frac{|c_l|^2}{2}-
\frac{3}{4}|c_l|^4-\frac{\nu^2}{4}+|c_l|^2\sqrt{\left(1-|c_l|^2
\right)^2-\nu^2}\cos\Phi\right)\nonumber
\label{efunc}
\end{eqnarray}

\noindent where $\Phi=2\alpha_l-\alpha_{l+1}-\alpha_{l-1}$ and $g$ is
the dimensionless interaction strength

\[
g\equiv(N-1)\frac{2MR^2}{\hbar^2}\Lambda=(N-1)\frac{4Ra}{S}.
\]

\noindent Note the $(N-1)$ factor included in this constant. This is
done so that, if the number of particles $N$ is increased {\it at
  fixed $g$}, (e.g. by scaling $a$ with $1/(N-1)$ and keeping the
geometrical parameters $R$ and $S$ fixed), then the one- and two-body
contributions to the energy functional will both scale with
$N$\cite{liebook}. The fact that the energy functional depends on the
phases of the variational parameters $c_j$ only through the
combination $\Phi$ is a consequence of invariance under the
transformation $\alpha_{l+k}\rightarrow\alpha_{l+k}+k\phi$,
$k=0,\pm1$, associated with the rotational invariance of $H$,
together with invariance under a global phase change
$\alpha_j\rightarrow\alpha_j +\phi$, associated with the global gauge
invariance of $H$. Note also that the functional (\ref{efunc})
satisfies the Bloch decomposition (\ref{blodec}), being of the form

\begin{equation}
{\cal{F}}_l(|c_l|,\Phi,\nu)=\frac{L^2}{N^2}+
{\cal{F}}_l^{\rm int}(|c_l|,\Phi,\nu),\hspace{1cm}L=N(l+\nu).
\label{vyrast}
\end{equation}

\noindent The last term of (\ref{vyrast}) is independent of $l$,
implying the periodicity ${\cal{F}}_l^{\rm int}
(|c_l|,\Phi,\nu)={\cal{F}}_0^{\rm int}(|c_0|,\Phi,\nu)$. This allows
in particular to restrict further considerations to $l=0$ only in
connection with the intrinsic part of the energy functional.

\begin{figure}
\centering
\includegraphics[width=3.6in]{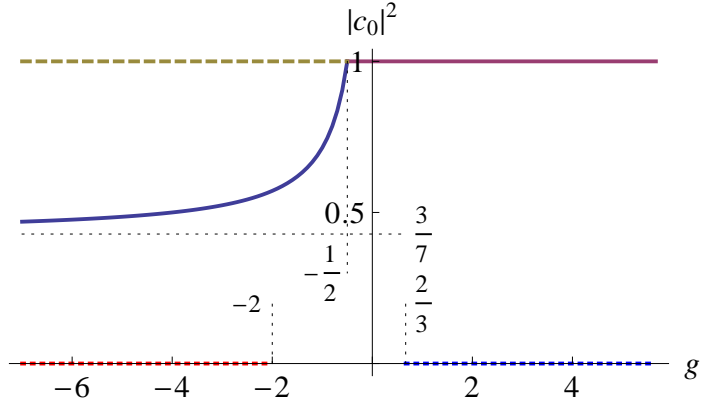}
\renewcommand{\figurename}{Fig.}
\caption{\small Chart of the relevant minima of the intrinsic energy
  functional ${\cal{F}}_l^{\rm int}(|\bar{c_0}|,\bar{\Phi},\nu)$ in
  the $|c_0|^2\times g$ plane for $\nu=0$. The curve at $|c_0|^2>
  \frac{3}{7}$, $g<-0.5$ corresponds to absolute minima with
  $\Phi=0$. It joins the straight line at $|c_0|^2=1$ which
  corresponds to `infima' with $\Phi=0,\pi$ for $g>-0.5$. Secondary
  minima are indicated at $|c_0|^2=0$, with $\Phi=\pi$ for $g<-2$ and
  $\Phi=0$ for $g>\frac{2}{3}$, and secondary `infima' occur at
  $|c_0|^2=1$ with $\Phi=\pi$ and $g<-0.5$.}
\label{funcland}
\end{figure}

The global minima of the intrinsic energy functional are taken as
variational approximations to the intrinsic yrast energies, i. e.

\[
\left.\begin{array}{l}
\frac{\partial{\cal{F}}_0^{\rm int}}{\partial\Phi}(|\bar{c_0}|,
\bar{\Phi},\nu)=0\\ \\
\frac{\partial{\cal{F}}_0^{\rm int}}{\partial|c_0|}(|\bar{c_0}|,
\bar{\Phi},\nu)=0 \end{array}\right\}
\;\,\Rightarrow\;\,e_{\rm yr}^{\rm int}(\nu)={\cal{F}}_0^{\rm int}
(|\bar{c_0}(\nu)|,\bar{\Phi},\nu) 
\]

\noindent An exception occurs for $\nu=0,\;\,g>-0.5$, in which case
the minimum reduces to an `infimum' at $|c_0|=\lim_{\nu\rightarrow 0+}
|\bar{c_0}(\nu)|=1$. This infimum appears in this domain of the
coupling constant $g$ as the $\nu\rightarrow 0$ limit of actual
absolute minima of the energy functional for non vanishing values of
$nu$. This behavior leads in fact to a cusp behavior of 
$e_{\rm yr}^{\rm int}(\nu)$ at $\nu=0$.

\begin{figure}
\centering
\includegraphics[width=3.2in]{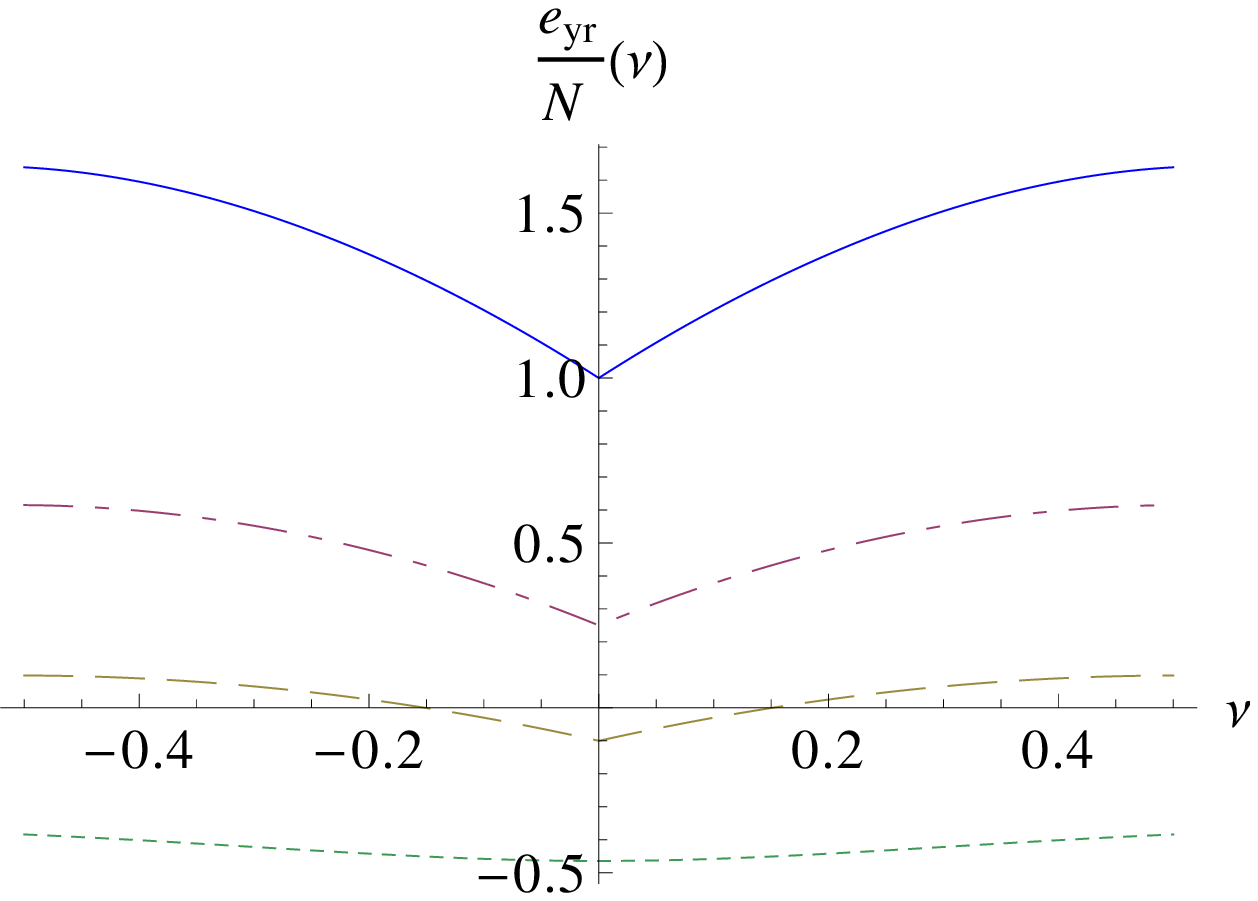}
\includegraphics[width=3.2in]{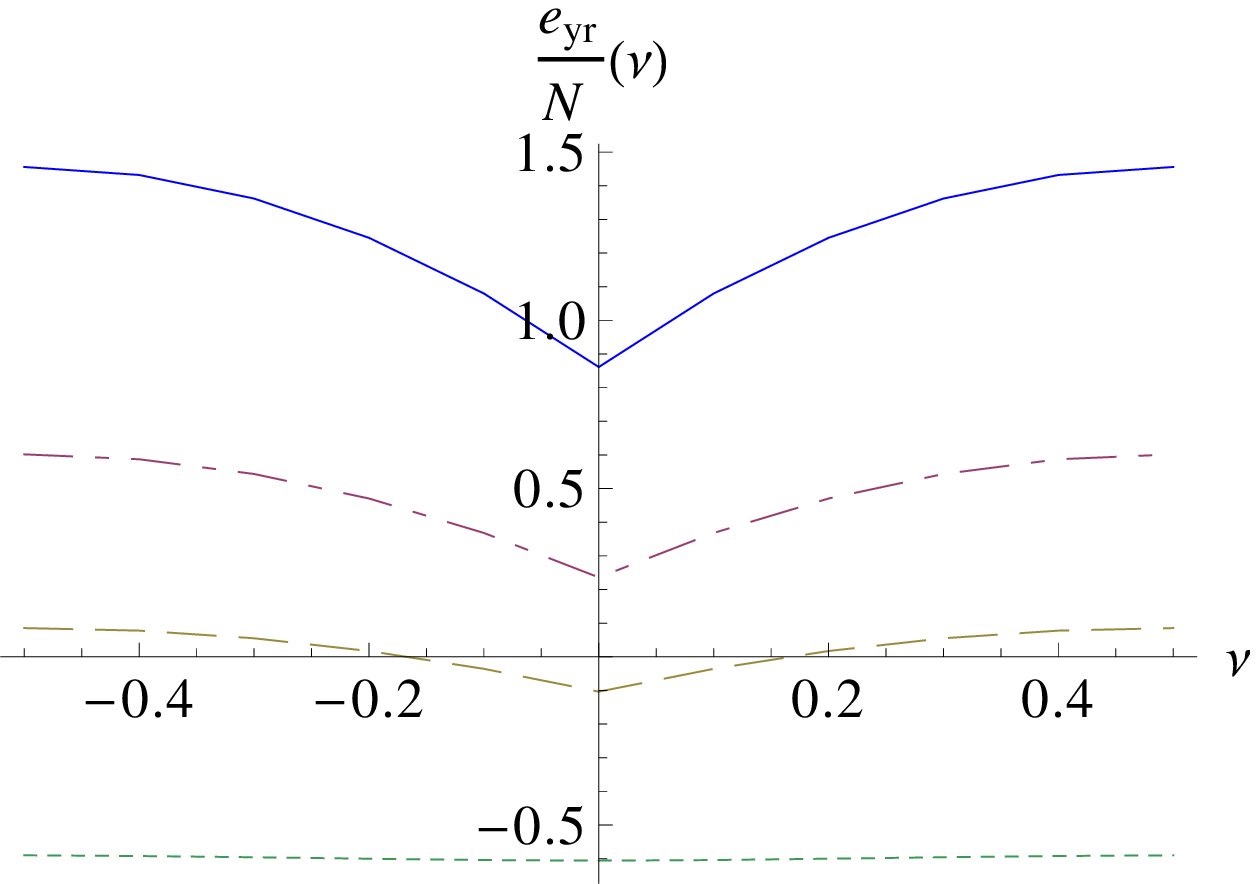}
\renewcommand{\figurename}{Fig.}
\caption{\small {\bf Top:} variational intrinsic yrast energies per
  particle (in units of $\hbar^2/2MR^2$) as a function of
  $\nu=(L-Nl)/N$ for $g=2$ (full line), $g=0.5$ (dot-dashed), $g=-0.2$
  (dashed) and g=-0.8 (dotted). The cusp minimum at $\nu=0$ is typical
  for $g>-0.5$. Extending the plot to $\nu>0.5$ (or to $\nu<-0.5$) by
  using periodicity generates a cusp maximum at half-integer values of
  $\nu$. This results from deterioration of the adopted
  variational ansatz away from the integer values of the abscissa. {\bf
  Bottom:} results obtained by numerical diagonalization, with $N=10$,
  $l_b=3$ and $m_0=0$, for the same quantities and for the same
  values of $g$ on the left hand side.} 
\label{yrvar}
\end{figure}

A chart of the relevant minima of the intrinsic energy functional is
shown in fig. \ref{funcland}, and numerical results for the
variational intrinsic yrast energies are displayed in
fig. \ref{yrvar}. For values of the coupling constant $g$ greater than
-0.5 one gets the cusp minimum at $\nu=0$, which translates also to
all integer values of $L/N$, at which the minimizer condensate
$|\Psi\rangle$ has good total angular momentum $L=Nl$. For values of
$g$ smaller than $-0.5$, the absolute minima at $\nu=0$ have
$|c_0|^2<1$ implying that in this domain one has $|c_{\pm 1}|\neq 0$,
so that the minimizer breaks the rotational symmetry. This signals a
quantum phase transition at $g=-0.5$, as pointed out in
ref. \cite{Rina} in connection with the ground ($L=0$)
state. Furthermore, as $g$ is decreased below the transition value,
calculated yrast intrinsic energies become very weakly dependent on
$\nu$. Consequently, the yrast line essentially reduces in this domain
of values of $g$ to a rotational band with the rigid moment of
inertia. Further insight into the yrast states in this phase will be
drawn from results of the numerical diagonalization of the model
Hamiltonian (\ref{Hbase}), to be discussed below.

\subsection{Persistent currents.}

\begin{figure}
\centering
\includegraphics[width=3.5in]{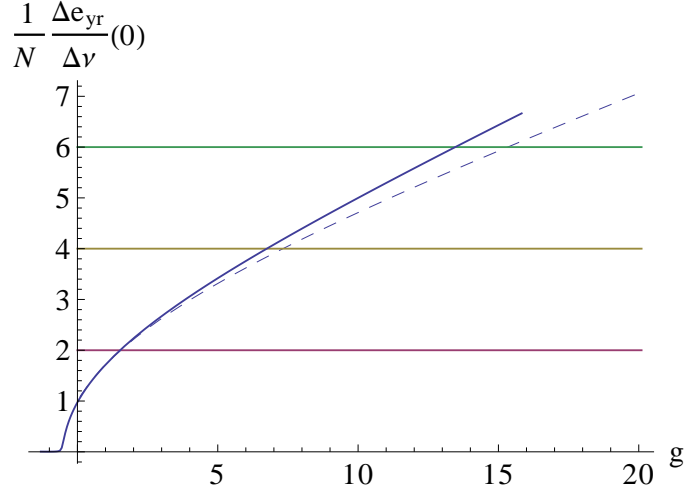}
\renewcommand{\figurename}{Fig.}
\caption{\small The {\bf dashed} curve shows the value of the slope
  $\frac{1}{N}\frac{\Delta e_{yr}}{\Delta\nu}$ at $\nu=0$ (in units of
  $\hbar^2/2MR^2$) as function of $g$, and the horizontal lines show
  the values of the slope of the center of mass energy per particle
  for $L/N=1,2$ and $3$. In view of the periodicity and symmetry
  properties of the intrinsic energies, the intercepts indicate the
  values of $g$ above which cusp minima develop at these values of
  $L/N$, indicating meta-stability of the currents in the sense
  proposed by Bloch\cite{Bloch}. The {\bf full line} curve shows the
  corresponding results obtained by diagonalizing the Hamiltonian
  (\ref{Hbase}) for $N=40$, $l=0$ and $l_b=2$ (5 single-particle
  states). The variational slopes have been obtained numerically using
  an increment $\Delta\nu=1/40$ to conform to the discreteness in the
  diagonalization calculation.}
\label{metavar}
\end{figure}

It was argued by Bloch\cite{Bloch} that the occurrence of minima at
non vanishing values of the angular momentum along the yrast line
indicated meta-stability of the associated current against additional
couplings allowing for exchange of angular momentum between the system
and its surroundings. The above variational results show that such
minima will in fact develop at integer values of $L/N$, albeit as cusp
minima, whenever the coupling parameter $g$ is large enough to make
the right-slope of the cusps shown in fig. \ref{yrvar} larger than the
slope of the center of mass energy at the corresponding value of
$L/N$. Having in mind the periodicity and symmetry properties of the
intrinsic yrast energies, it is in fact easy to see that cusp minima
will in this case appear in the yrast line itself. Quantitative
predictions of the values of $g$ at which this happens for the three
lowest integer values of $L/N$ are shown in fig. \ref{metavar}.

It is worth noting explicitly that the meta-stability criterion, based
on the persistence of local yrast minima in the present model,
actually refers to what may be called `one-dimensional
meta-stability', as it depends in an essential way on the assumption
that the transverse degrees of freedom in the quasi-one-dimensional
system are effectively frozen. However, a simple qualitative
observation which can be made concerning the $L$-values at which
instabilities appear to become an experimentally limiting factor for
the observation of persistent currents\cite{exphslip,expLargeL} is
that they occur where the de Broglie wavelength is no longer much
larger than the transverse scale of the employed toroidal trap, which
would allow for the participation of additional degrees of freedom in
the dynamics of angular momentum transfer. Also, in these experiments,
the values of the coupling constant $g$ can be estimated to be of the
order of thousands, indicating persistence of `one-dimensional
meta-stability' for values of $L$ up to about two orders of magnitude
above the largest reported observed value. A realistic appraisal of
the decay of the macroscopically quantized angular momentum of the
persistent currents is therefore beyond the bearings of the present
model.

\section{Characterization of the yrast line II - numerical
  diagonalization.}

For not too large a number of particles the stationary states of the
Hamiltonian (\ref{Hbase}) can be obtained by direct numerical
diagonalization in a truncated base of many-body state vectors. Taking
advantage of the fact that the total angular momentum is a constant of
motion, one may consider for each value of $L$ the truncated base
formed by the set of $N$-body state vectors

\begin{equation}
{\prod_m}'\frac{1}{\sqrt{m!}}(b_m^\dagger)^{n_m}|0\rangle
\label{mbbase}
\end{equation}

\noindent with $l-l_b\leq m\leq l+l_b$, $\sum_m n_m=N$ and $\sum_m
mn_m=L$. The value of $l$ plays here the same role as in
eq. (\ref{constr1}), i.e. shifting the central value of the selected
band of $2l_b+1$ single particle angular momentum functions in order
to optimize the base for the relevant value of $L$, taking into
account the periodicity and symmetry properties of the intrinsic
spectrum. The $n_m$ are occupation numbers for the single particle
states $m$. We report on calculations involving values of $N$ from 10
to 40, and bases involving from 5 $(l_b=2)$ to 11 $(l_b=5)$
single-particle states. Due to the periodicity of the (non-trivial)
intrinsic spectrum, we set $l=0$ without loss of generality. It is
worth noting that, for a given number of particles $N$, the truncation
involved in the adoption of a given set of single particle states
leads to many-body subspaces with good $L$ having different
dimensionalities, implying a small residual $L$ dependence of the
truncation effects on the many-body eigenfunctions and energy
eigenvalues. The latter are in any case variational upper bounds to
the `true' eigenvalues for each value of $L$.

\begin{figure}
\centering
\includegraphics[width=3.5in]{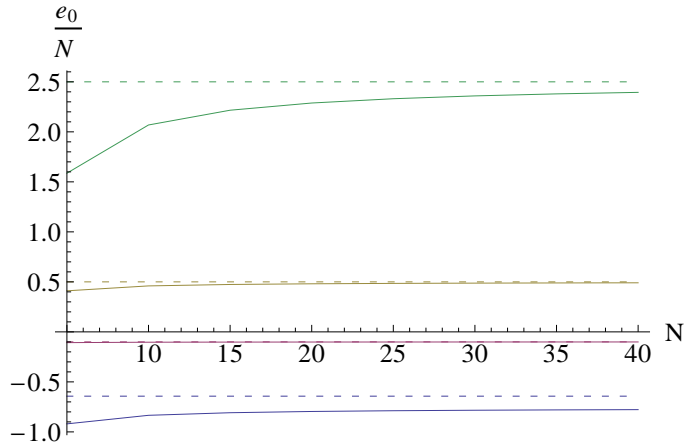}
\renewcommand{\figurename}{Fig.}
\caption{\small Variational $L=0$ ground state energies per particle
  for $g=-1.0,\;-0.2,\;1.0$ and $5.0$ (dashed lines, from the bottom
  up) and the corresponding $L=0$ ground state energies per particle
  obtained by numerical diagonalization for a range of values of
  $N$ using the prescription (\ref{Lamg}). The many-body bases are
  constructed from a set of single particle orbitals with $l=0$ and
  $l_b=2$ (see (\ref{mbbase})).} 
\label{Nscal}
\end{figure}

The value of the effective two-body interaction strength parameter
$\Lambda$ is always taken, in the numerical diagonalizations, to be
related to the strength parameter $g$ used in connection with the
variational treatment as

\begin{equation}
\Lambda=\frac{\hbar^2}{2MR^2}\frac{g}{N-1}
\label{Lamg}
\end{equation}

\noindent with the understanding that different particle numbers $N$
are dealt with by adjusting $\Lambda$ so as to keep the value of $g$
constant, as this will tend to preserve the `ground state phase
diagram' of the system as $N$ is varied. This is indeed an {\it exact}
feature of the variational approach, but holds only approximately when
one carries out many-body diagonalizations, as shown explicitly in
fig. \ref{Nscal}, which compares the variational energy per particle
for the $L=0$ ground state evaluated variationally for some sample
values of $g$ and the corresponding $N$-dependent energies per
particle obtained by diagonalization in many-body bases built on a set
of single-particle orbitals with $l=0$ and $l_b=2$.

Intrinsic yrast energies obtained in this way are shown in
fig. \ref{yrvar} together with the corresponding variational results,
showing the very good overall agreement of the two evaluations, even
though the diagonalization is performed for a small system ($N=10$).
It should be kept in mind that, in view of the truncation of the base
the reported yrast energy results actually provide upper bounds for
the `actual' yrast energies; and that, for results obtained using
many-body bases constructed from a fixed set of single-particle
orbitals, truncation effects are apt to increase as one moves away
from the cusp at $\nu=0$. In particular, residual $L-$dependence of
the intrinsic yrast energies $e_{\rm yr}$ in the deformed domain
$g<-0.5$ is reduced as $l_b$ (i.e., the number of single-particle
orbitals involved in the construction of the many-body base vectors)
is increased.

The predicted values of $g$ at which the cusps generate yrast minima
indicating the onset of meta-stable currents at integer values of
$L/N$ can be read from the values of $de_{\rm yr}(0)/d\nu$ shown as
the full line curve of fig. \ref{metavar}. Comparison with the values
obtained from the variational dashed curve reveal that the truncated
diagonalization results in deeper cusp minima, leading to smaller
critical values of $g$ notably for $L/N\geq 2.$

\section{Probing correlated states.}

The availability of correlated many-body yrast state vectors obtained
from many-body diagonalizations allows for further probing of the
variational transition at $g=-0.5$. As found also in the variational
calculation, numerical diagonalizations in this domain reveal a regime
in which the intrinsic yrast energies become essentially independent
of $\nu=L/N-l$, the yrast line approaching a rotational band whose
moment of inertia is essentially the rigid moment of inertia. Note
that the intrinsic energies for different values of $\nu$ correspond
now to diagonalizations carried out in orthogonal many-body subspaces,
dynamically decoupled by symmetry. This suggests that the quantum
phase transition identified through the breaking of the rotational
symmetry of the variational ground state is akin to the onset of
`intrinsic deformations' in nuclear and molecular spectra displaying
well developed rotational bands.

\begin{figure}
\centering
\includegraphics[width=3.5in]{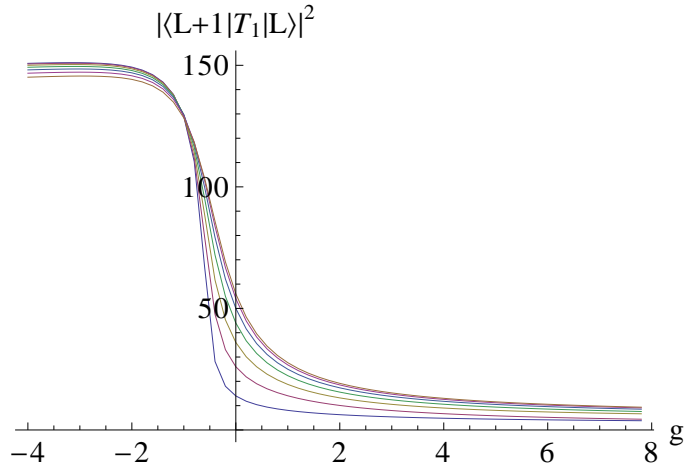}
\renewcommand{\figurename}{Fig.}
\caption{\small Angular momentum transfer matrix elements between
  yrast states with successive angular momentum values as a function
  of the coupling strength $g$ for $N=14$. The different curves
  correspond to different values of $L$. The value of the matrix
  elements for non interacting particles $(g=0)$ comes from bosonic
  enhancement factors and are respectively 14, 26, 36, 44, 50 54 and
  56 for $L=$ 0 to 6. The onset of the deformed phase for $g<-0.5$
  is accompanied by a large enhancement of these matrix elements.}
\label{trans}
\end{figure}

As one way of scrutinizing this interpretation, one may take advantage
of the good angular momentum eigenstates resulting from the numerical
diagonalization procedure in order to detect collective behavior in
appropriate transition matrix elements\cite{BM}. A simple and
effective choice in this connection is to evaluate matrix elements
between successive yrast states of the angular momentum transfer
operator

\[
T_1\equiv\sum_{m=l-l_b}^{l+l_b-1}\left(b^\dagger_{m+1}b_m+b_m^\dagger
  b_{m+1} \right).
\]

\noindent Matrix elements of this operator between yrast states for a
range of values of the coupling strength parameter $g$ are shown in
Fig. \ref{trans} for states with particle number $N=14$. A strong
enhancement above the values involving just bosonic factors (which
alone determine the matrix elements at $g=0$) sets in sharply in the
deformed domain $g<-0.5$. For $g>0$, on the other hand, the free
particle values are quenched by the correlations induced by the
effective two-body interaction. To the extent that operators involving
$T_1$ are involved in the loss of angular momentum trough external
couplings, this provides an indication that this type of loss is
hindered by the correlations induced by repulsive two-body
interactions.

\begin{figure}
\centering
\includegraphics[width=3.6in]{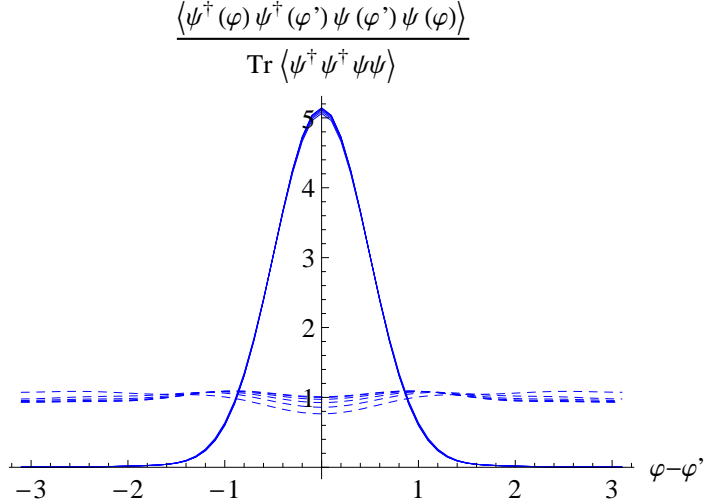}
\renewcommand{\figurename}{Fig.}
\caption{\small {\bf Dashed curves:} Nearly isotropic two-body
  correlation functions at $g=1$ for the yrast states with $L=0$ to
  $5$ (lowest to upper curve at $\varphi-\varphi'=0$), using many-body
  states obtained from a diagonalization of the Hamiltonian
  (\ref{Hbase}) with $N=10$, $l=0$, $l_b=4$ (see (\ref{mbbase})). {\bf
    Full curves:} Same as the dashed curves but at $g=-1.$ Note in
  this case the strong anisotropy and the weaker angular momentum
  dependence of the correlation functions.}
\label{corfun}
\end{figure}

As a second approach, one may use the numerical correlated eigenstates
to evaluate appropriate correlation functions. We focus in particular
on the correlation functions

\begin{equation}
\frac{\langle\Psi_{E_L}|\psi^\dagger(\varphi)\psi^\dagger(\varphi')
\psi(\varphi')\psi(\varphi)|\Psi_{E_L}\rangle}{{\rm Tr}\langle
\Psi_{E_L}|\psi^\dagger\psi^\dagger\psi\psi|\Psi_{E_L}\rangle}\equiv
\rho^{(2)}_{E_L}(\varphi- \varphi')
\label{ddcor}
\end{equation}

\noindent where the states $|\Psi_{E_L}\rangle$ are yrast
states. Results for this function are shown in fig. \ref{corfun} for
$g=1.$ and for $g=-1.$ While in the first case there is a clear
dependence on the value of the angular momentum (together with some
tendency to anti-correlation, as signaled by the relatively shallow
minima at $\varphi-\varphi'=0$), in the attractive case the
correlation function is essentially independent on the value of the
angular momentum, being strongly peaked at $\varphi-\varphi'=0$ and
dropping to very small values at $\varphi-\varphi'=\pm\pi$
consistently with the correlation expected as a result of the
attractive character of the effective interaction.

\begin{figure}
\centering
\includegraphics[width=3.2in]{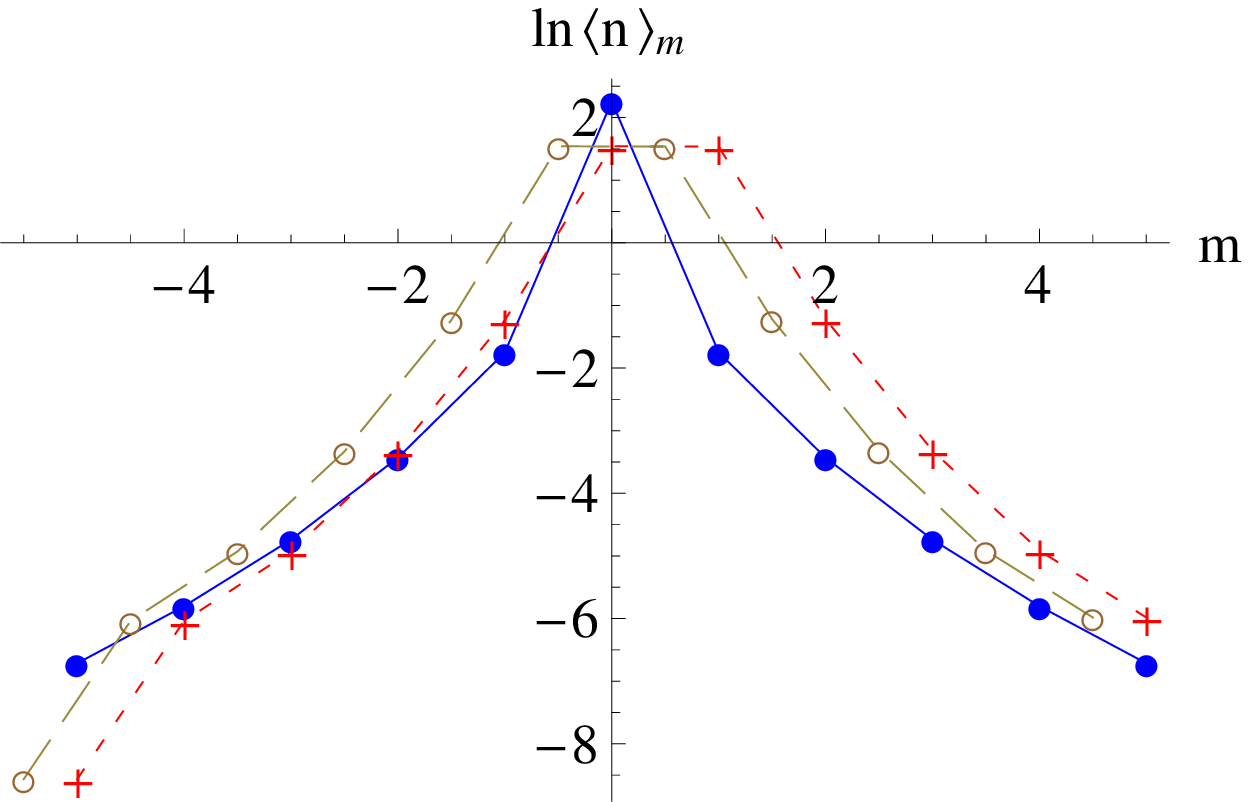}
\includegraphics[width=3.2in]{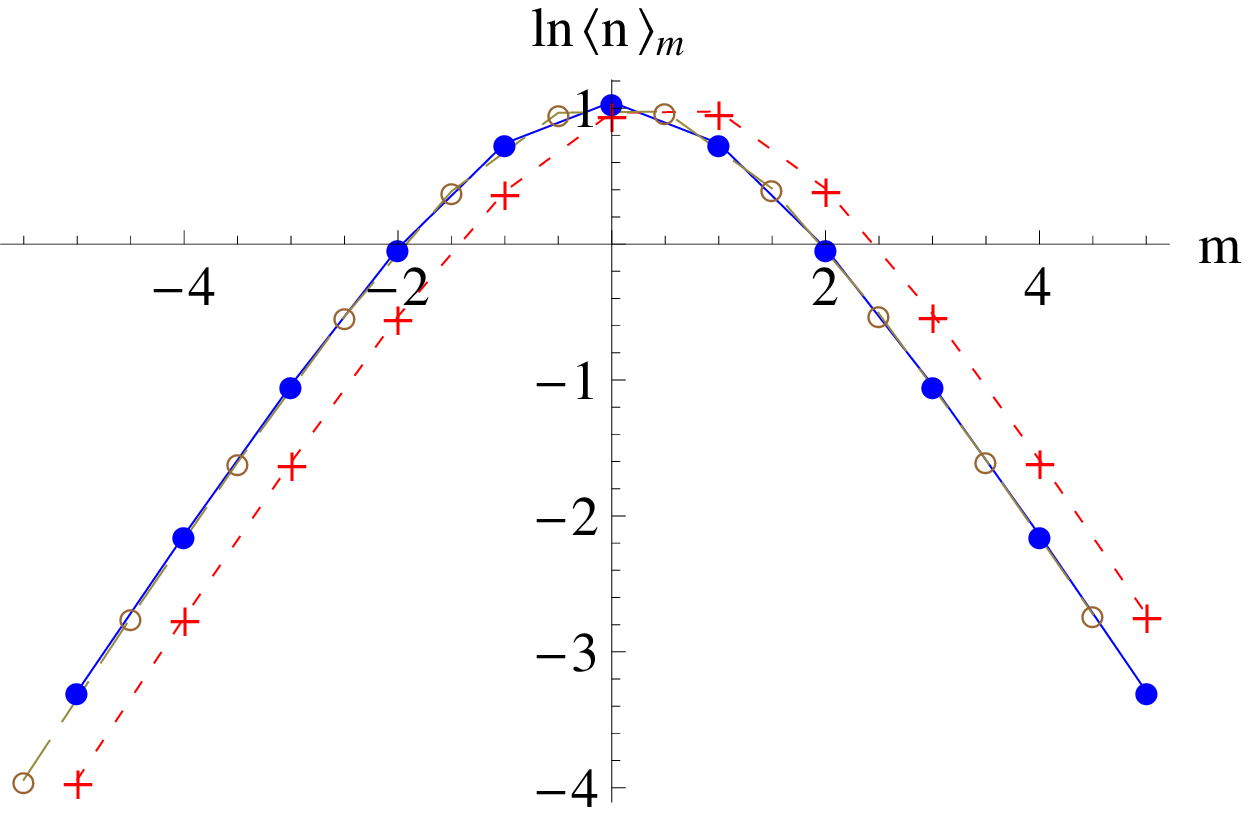}
\renewcommand{\figurename}{Fig.}
\caption{\small Logarithm (for clarity in the display of smaller
  values) of the mean single particle occupation numbers $\langle
  n\rangle_m$, for a system with $N=10$ treated in the many-body base
  constructed from single-particle orbitals with $l=0$ and $l_b=5$
  (see (\ref{mbbase})), plotted against the single particle angular
  momentum $m$. The plot on {\bf top} has been calculated with a
  repulsive effective interaction $g=2.0$ and displays results for the
  $L=0$ (filled circles, full line) and $L=5$ ($L/N=0.5$, crosses,
  short-dashed line). The open circles connected by a long-dashed line
  are those corresponding to $L/N=0.5$ displaced by $-0.5$ along the
  $m$ axis. The {\bf bottom} graph shows the corresponding plots for
  an attractive effective interaction with $g=-2.0$. Unlike in the case
  of the repulsive interaction, here the points obtained for
  $L/N=0.5$, after translation by $-0.5$ along the $m$ axis,
  essentially fall on the same curve as the $L=0$ occupations.}
\label{logone}
\end{figure}

The stiffness of the two-body correlation function with respect to
changes of angular momentum in the `deformed' phase in fact leaves an
imprint also in the one-body reduced density matrix

\[
\langle\Psi_{E_L}|b^\dagger_{m'}b_m|\Psi_{E_L}\rangle=
\rho^{(1)}_{E_L mm'}=\langle n\rangle_m\;\delta_{mm'}
\]

\noindent whose eigenvalues, written as $\langle n\rangle_m$, are the
mean occupation numbers of the single particle orbitals
$\phi_m(\varphi)$. As illustrated in fig. \ref{logone}, in the
deformed regime the eigenvalues of the one body density for yrast
states with different angular momenta fall essentially on a single
distribution curve, after being translated by the fractional value of
$L/N$. This is at variance with what one finds in the non-deformed
regime, where a considerable fraction of the particle number $N$ is
carried by the $m=0$ orbital consistently with the relatively small
anisotropy of the two-body correlation function in this case.

\subsection{Yrast many-body correlations.}

Finally, one may consider the full `intrinsic wave functions'
$\chi_{N,L}(\varphi_1, \dots,\varphi_N)$ defined by Bloch by factoring
out from the yrast wave functions $\Psi_{N,L}$ `center of mass factors'
for the given number of particles $N$ and for each value of $L$, i. e.

\[
\Psi_{N,L}(\varphi_1,\dots,\varphi_N)=
e^{i\frac{L}{N}\sum_{j=1}^N\varphi_j}\chi_{N,L}(\varphi_1,\dots,\varphi_N)
\]

\noindent These intrinsic wave functions depend only on coordinate
differences $\varphi_j-\varphi_k$ and are therefore rotationally
invariant. They satisfy the `twisted' boundary conditions

\[
\chi_{N,L}(\varphi_1,\dots,\varphi_j+2\pi,\dots,\varphi_N)=
e^{-2\pi i \frac{L}{N}}\chi_{N,L}(\varphi_1,\dots,\varphi_j
,\dots,\varphi_N).
\]

\noindent Since they are rotationally invariant, the intrinsic
wave functions are eigenfunctions of the total angular momentum with
eigenvalue zero, so that they are also eigenfunctions of $H$ with
eigenvalue $e_{\rm int}^{(L)}$.

In order to access the intrinsic wave functions for each of the values
of the total angular momentum $L$ within the second-quantization
framework adopted here we first consider the description of the system
in a frame of reference which rotates with angular velocity $\Omega$
about the axis of the toroidal configuration space. The corresponding
Hamiltonian $H_\Omega$ is related to $H$, as written in
eq. (\ref{Hfield}), by making the replacement $l_z^2\rightarrow
(l_z-MR²\Omega/\hbar)^2$ in the kinetic energy term. The terms
involving $\Omega$ may furthermore be eliminated by means the gauge
transformation of the field operators $\psi(\varphi)\equiv
e^{i\frac{MR^2\Omega}{\hbar}\varphi}\psi'(\varphi)$, which leads to
the simpler form of $H_\Omega$ involving the `twisted' field operators
$\psi'(\varphi)$

\[
H_\Omega= \frac{\hbar^{2}}{2 M R^{2}} \int_{0}^{2 \pi} d\varphi
\psi'^{\,\dagger}(\varphi) l_{z}^{2} \psi'(\varphi)++\frac{1}{2}
\frac{U_{c}}{RS}\int_{0}^{2 \pi} \psi'^{\,\dagger} (\varphi)
\psi'^{\,\dagger}(\varphi)\psi'(\varphi) \psi'(\varphi) d\varphi.
\]

\noindent Consider next the representation of this Hamiltonian in the
orthonormal twisted single-particle base with vectors

\begin{equation}
\phi_{m}^{\Omega}(\varphi)=\frac{1}{\sqrt{2 \pi}} e^{i\left(m-
\frac{MR^2\Omega}{\hbar}\right)\varphi}\;,\hspace{0.5cm}
\frac{1}{2\pi}\int_0^{2\pi}d\varphi\;\phi_{m'}^{\Omega\,*}(\varphi)
\phi_{m}^{\Omega}(\varphi)=\delta_{mm'}
\label{twstbase}
\end{equation}

\noindent by expanding the field operators as

\begin{equation}
\psi'(\varphi)=\sum_m\phi_{m}^{\Omega}(\varphi)\beta^{\Omega}_m.
\label{twist}
\end{equation}

\noindent The resulting expression for the second quantized
Hamiltonian in the rotating frame is therefore

\begin{eqnarray*}
H_\Omega&=&\frac{\hbar^{2}}{2 M R^{2}}\sum_{m}\left(m-\frac{MR^2\Omega}
{\hbar}\right)\beta^{\Omega\,\dagger}_{m}\beta^\Omega_{m}+ \\
&&\hspace{2cm}
+\frac{\Lambda}{2}\sum_{m_i}\beta^{\Omega\,\dagger}_{m_{1}}
\beta^{\Omega\,\dagger}_{m_{2}}\beta^\Omega_{m_{3}}\beta^\Omega_{m_{4}}
\delta_{m_{1}+m_{2},m_{3}+m_{4}}.
\end{eqnarray*}

\noindent Particle number and the total angular momentum remain
constants of motion.

When considering states with with $N$ particles and total angular
momentum $L$, the choice $\Omega\rightarrow\Omega_{NL} \equiv\hbar
L/(NMR^2)$ reduces the gauge term to just $L/N$. With this choice, in
the $N$, $L$ sector of the Fock space the Hamiltonian $H_\Omega$
reduces to to the form $(\ref{Hbase})$ with the $b$-operators replaced
by twisted $\beta$-operators, and with the subtraction of the center
of mass energy $\hbar^2L^2/(2NMR^2)$. Thus, except for the trivial
subtraction of a multiple of the unit matrix related to the center of
mass energy, it is represented in the basis of twisted many-body state
vectors vectors (cf. eq. (\ref{mbbase}))

\[
\prod_m\frac{1}{\sqrt{m}}(\beta_m^{N,L\;\dagger})^{n_m}|0\rangle,
\hspace{.5cm}\sum_mn_m=N,\;\,\sum_mmn_m=L
\]

\noindent by the same numerical matrix obtained in connection with the
Hamiltonian (\ref{Hbase}) using the many-body states (\ref{mbbase}).

This identifies the numerical ground state in the {\it twisted} $N$,
$L$ sector as the yrast `intrinsic state' for the selected values of
$N$ and $L$. The `intrinsic states' are therefore just of the form
(\ref{mbstate}), with the same coefficients $C^{(j)}_{\{n_m\}_{N,L}}$,
but with the $N$-body base vectors expressed in terms of the `twisted'
bosonic operators $\beta^{\Omega_{NL}}_m$,
$\beta^{\Omega_{NL}\,\dagger}_m$, i.e.

\[
|\Psi^{\rm int}_{e^{(L)}_{\rm int}}\rangle=\sum_{\{n_m\}_{N,L}}
C^{(0)}_{\{n_m\}_{N,L}}\prod_m\frac{1}{\sqrt{n_m!}}\left(
\beta^{\Omega_{NL}\,\dagger}_m\right)^{n_m}|0\rangle.
\]

\noindent The fact that they correspond to rotationally invariant
amplitudes in configuration space is borne out by the fact that
$\sum_mn_m(m-L/N)=0$. It should be stressed that intrinsic states
corresponding to different values of the total angular momentum $L$
will {\it not} be orthogonal, since they are eigenvectors of
differently twisted Hamiltonians.

\begin{figure}
\centering
\includegraphics[width=3.2in]{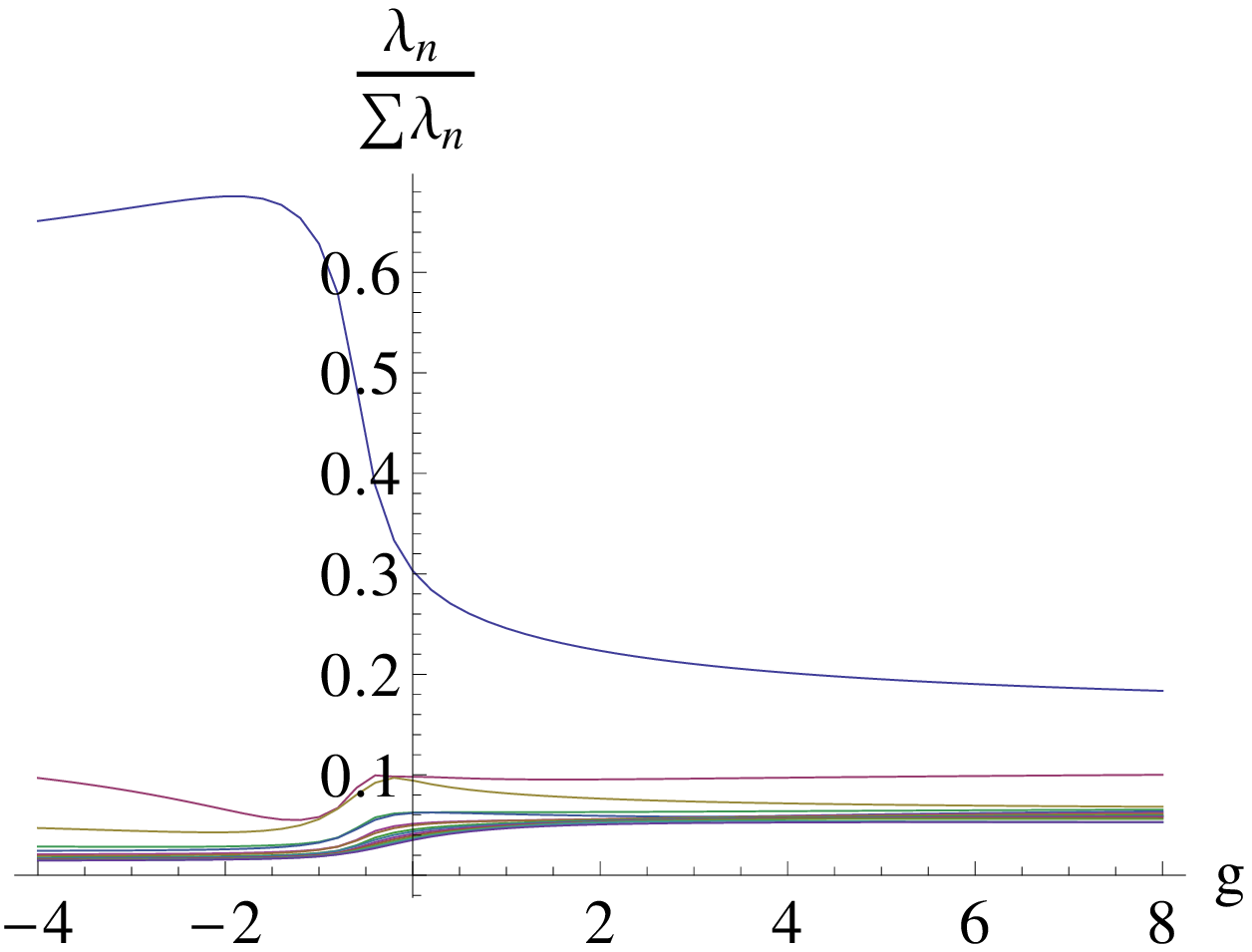}
\includegraphics[width=3.5in]{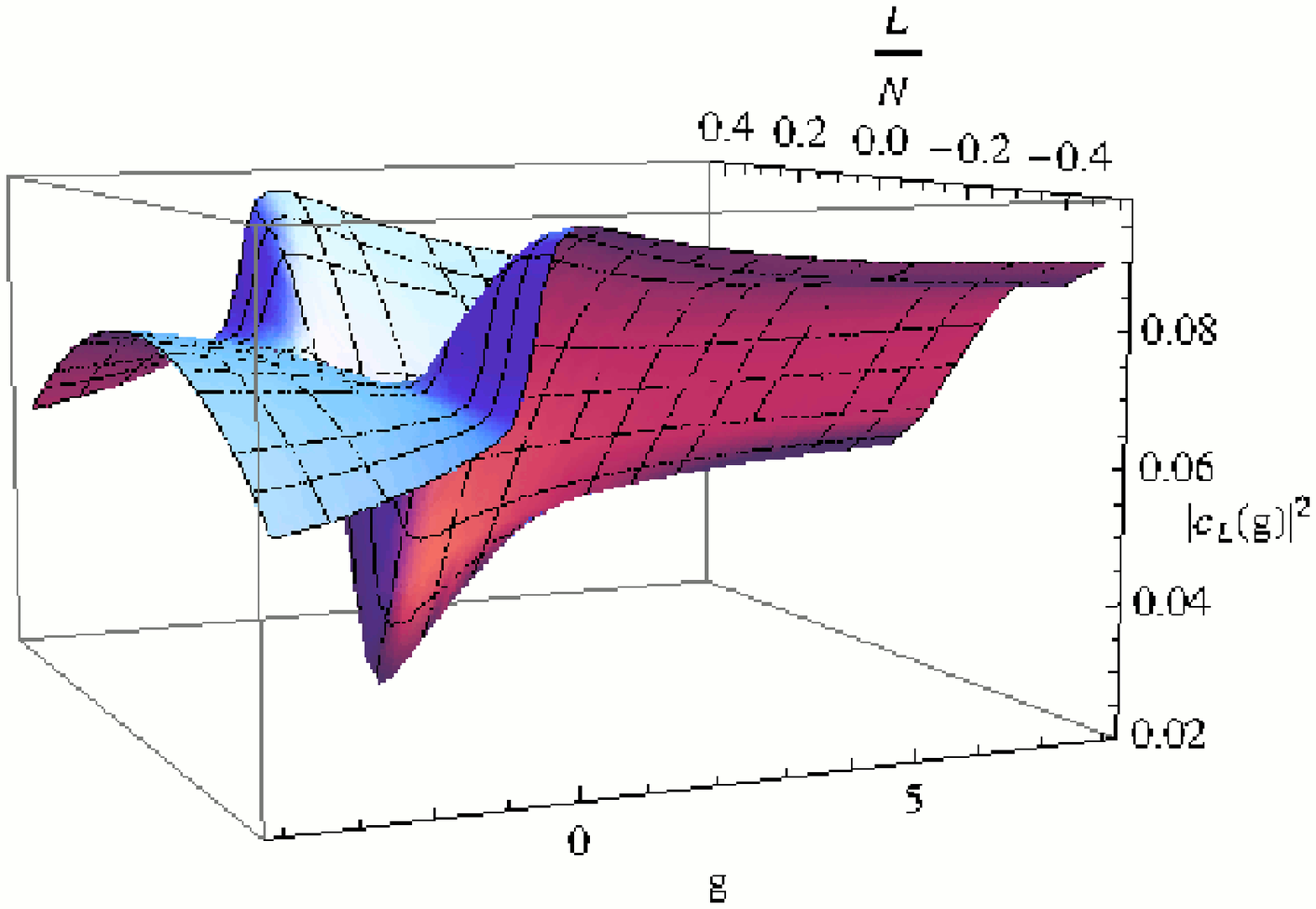}
\renewcommand{\figurename}{Fig.}
\caption{\small {\bf Top:} Eigenvalues $\lambda_n$ of the overlap
  matrix of many-body intrinsic states for $N=14$, $L=-7$ to $6$ as
  functions of the coupling strength $g$. The values plotted are
  normalized by the trace ($\sum\lambda_n=14$) of the overlap matrix.
  {\bf Bottom:} Squared amplitudes along intrinsic states with
  different $L$-values of the eigenvector associated with the largest
  eigenvalue of the overlap matrix as functions of $g$. The numerical
  many-body eigenstates used to generate the overlap matrix were
  evaluated using the full many-body base constructed in terms of five
  single-particle orbitals with $l=0$ and $l_b=2$ (see
  (\ref{mbbase})).}
\label{ovlmat}
\end{figure}

These rotationally invariant intrinsic yrast states can be used, in
particular, to provide for indications concerning the nature of the
quantum phase transition at $g\simeq-0,5$. The simplest information
which may be derived from these non-orthogonal states is the content
and hierarchy of the domain of the many-body phase space spanned by
them. For the purpose of obtaining it we take the set of yrast
intrinsic states associated with one period, i. e. $-\frac{1}{2}\leq
\frac{L}{N}<\frac{1}{2}$, and evaluate their overlap matrix

\begin{equation}
\Omega^{(N)}_{L,L'}=\langle\Psi^{\rm int}_{e^{(L)}_{\rm int}}|
\Psi^{\rm int}_{e^{(L')}_{\rm int}}\rangle.
\label{intovlp}
\end{equation}

\noindent This is done using the commutators

\[
\left[\beta^{N,L}_m,\beta^{N,L'\;\dagger}_{m'}\right]=\left(
\phi^{N,L'}_{m'}|\phi^{N,L}_m\right)\equiv\int_0^{2\pi}d\varphi
\phi^{N,L'\,*}_{m'}(\varphi)\phi^{N,L}_m(\varphi)
\]

\noindent which involve the unitary transformation between different
twisted single particle bases, these being defined in
eq. (\ref{twstbase}). Diagonalizing the overlap matrix, i.e.

\[
\sum_{L'}\Omega^{(N)}_{LL'}f^{(n)}_{L'}=\lambda_nf^{(n)}_{L},
\hspace{.5cm}{\rm with}\hspace{.5cm}\sum_Lf^{(n)*}_{L}f^{(n')}_{L}=\delta_{nn'}
\]

\noindent so that

\[
\Omega^{(N)}_{LL'}=\sum_n\lambda_nf^{(n)}_{L}f^{(n)*}_{L'},
\]

\noindent we see that the eigenvectors $f^{(n)}$ identify `principal
directions' in the many-body phase space, as their components
$f^{(n)}_{L}$ represent amplitudes of each principal direction on the
different intrinsic states. The associated eigenvalues
$\lambda_n$, on the other hand, specify weights identifying the
participation of each one of the principal directions in the set of
intrinsic states.

The effect of the onset of deformation on the set of intrinsic states
is borne out in terms of these ingredients in Fig. \ref{ovlmat}. The
upper part of this figure shows a plot of the relative weight of the
largest eigenvalue $\lambda_0$ of the overlap matrix as function of
the coupling constant $g$, while the lowest part plots squared
components $|f^{(0)}_{L}|^2$ of the eigenvector associated to the
largest eigenvalue along the intrinsic states. A clear transition is
seen to occur in the spectrum of the overlap matrix when the value of
the coupling constant is reduced in the vicinity of $g=-0.5$. In the
domain with stronger attractive coupling a single state acquires
dominance over the content of the intrinsic subspace, as revealed by
the fact that the corresponding eigenvalue exhausts an important
fraction of the trace. This indicates that the yrast states in the
`deformed phase', at $g<-0.5$, are at least strongly dominated by a
single, common many-body correlation function. A change of regime is
also noted in the $L$-composition of the dominant eigenvector, which
becomes uniformly flat in the `deformed' region, as shown also in
Fig. \ref{ovlmat}. This indicates that the dominant, common intrinsic
state is a nearly uniform superposition of the intrinsic states
for the various $L$-values present in the intrinsic yrast period.

The features shown in Fig. \ref{ovlmat} for $N=14$ particles are found
also for smaller values on $N$, albeit with a strengthening of the
single state dominance. This may be attributed at least in part to
limitations imposed by the truncation of the twisted bases on the
numerical evaluation of the many-body intrinsic overlaps. Indeed the
truncation is apt to fail to saturate the unitarity relations assumed
in the evaluation of the overlaps. Because of this, the calculated
value of the dominant eigenvalue should be considered as a lower
bound.

\section{Concluding remarks.}
  
We have examined a particular realization of the one-dimensional model
system considered long ago by F. Bloch\cite{Bloch}. It involves
ingredients drawn from current experimental and theoretical research
involving trapped, dilute atomic gases. In particular, the usual
effective contact two-body interaction based on the atom-atom
scattering length and its quasi one-dimensional reduction involving
the trap geometry have been explicitly used. The main focus is the
quantum dynamics of finite many-body systems, in which invariance
under rotations and global gauge transformations play a fundamental
role in establishing relevant dynamic cleavages between subspaces of
the many-body phase space with different values of total angular
momentum and particle number. In this context, the phenomena of
persistent, meta-stable currents and of `deformations', as revealed by
collective rotational features, appear as emergent properties
developing across the cleavages imposed by symmetry through the onset
of appropriate correlation properties. A natural domain hosting such
emergence is the set of yrast states, the lowest energy states for
each value of the total angular momentum.

The properties of the yrast states resulting from a variational
mean-field approach allowing for the control of the mean total angular
momentum, which amounts to a restricted mean field approximation, have
been confronted to correlated many-body stationary states resulting
from a many-body diagonalization in truncated Fock subspaces with good
particle number and angular momentum.

The effects of a sufficiently strong repulsive two-body effective
interactions appear in both cases as essential for the generation of
the yrast energy minima which can be associated with meta-stability of
persistent currents\cite{Bloch,leggett}. The relevant minima appear as
yrast cusps at values of the total angular momentum which are integer
multiples of the particle number. Threshold two-body interaction
values for the occurrence of actual minima at a given integer value of
the total angular momentum per particle become rather insensitive to
particle number $N$ when scaled by $1/N$, as this maintains the
balance of the kinetic and interaction effects in the effective
Hamiltonian.

In the case of attractive two-body effective interactions, and within
the variational restricted mean-field approach, the `quantum phase
transition' pointed out in ref. \cite{Rina} manifests itself, as
usual, in terms of a self-consistence related breaking of the
rotational symmetry of the ground state. This is moreover accompanied
by the disappearance of the yrast cusps, which are replaced by
analytic, locally quadratic yrast minima together with the overall
flattening of the intrinsic yrast energies.  This behavior of the
intrinsic yrast energies is borne out also in the case of the
diagonalization results. Since they rely on the use of subspaces of
good particle number and angular momentum, properties related to the
quantum phase transition must be sought in the correlation properties
of many-body states. In fact not only the behavior of two-body
correlation functions of the yrast states undergoes profound changes
as one crosses the transition region, but further standard
characteristics of collective `deformations'\cite{BM} also arise
there. In fact we find both a strong `collective' enhancement of
transition matrix elements of a one-body angular momentum transfer
operator acting between neighboring members of the yrast line and the
emergence of a weighty common eigen-component in the many-body
subspace generated by the intrinsic yrast states. The individual
intrinsic yrast states are themselves rather uniformly represented in
this common eigen-component, as can be seen in figure \ref{ovlmat}.
The full `deformation syndrome', present notably in the dynamics of
the much more complex molecular and nuclear systems, appears thus to
be deployed in the present one-dimensional model. The simplicity of
the model makes the associated correlation structures more accessible
to detailed scrutiny.

\vspace{.2cm}

\noindent {\bf Acknowledgement.} We acknowledge useful discussions
with Jo\~{a}o C. A. Barata.


\begin{thebibliography}{99}
\bibitem{note} While this obviously the case for the standard contact
  effective interaction adopted in eqs. (\ref{Hfield}) and
  (\ref{Hbase}), it remains true under the weaker condition of
  rotational invariance of the two-body interaction term\cite{Bloch}.
\end{thebibliography}
\end{document}